\documentclass[aps,prb,reprint,superscriptaddress]{revtex4-1}
\usepackage{amsmath}
\usepackage{multirow}
\usepackage{graphicx}
\usepackage{epstopdf}
\usepackage{tabularx, booktabs}
\usepackage{pbox}
\newcolumntype{Y}{>{\centering\arraybackslash}X}
\usepackage[usenames,dvipsnames]{color}
\usepackage[caption=false,position=b,singlelinecheck=off,font=normalsize,labelfont=bf,justification=justified]{subfig}
\usepackage{hyperref}

\hypersetup{
	colorlinks=true,
	linkcolor=blue,
	filecolor=black,      
	urlcolor=blue,
	citecolor=blue,
}
\graphicspath{ {Figures_PDF/} }

\begin{document}

\title{Persistence of antiferromagnetic order upon La substitution in the $\mathbf{4d^4}$ Mott insulator Ca$_2$RuO$_4$} 

\author{D. Pincini}
\email[]{davide.pincini.14@ucl.ac.uk}
\affiliation{London Centre for Nanotechnology and Department of Physics and Astronomy, University College London, Gower Street, London WC1E6BT, UK}
\affiliation{Diamond Light Source Ltd., Diamond House, Harwell Science \& Innovation Campus, Didcot, Oxfordshire OX11 0DE, UK}

\author{S. Boseggia}
\affiliation{London Centre for Nanotechnology and Department of Physics and Astronomy, University College London, Gower Street, London WC1E6BT, UK}

\author{R. Perry}
\affiliation{London Centre for Nanotechnology and Department of Physics and Astronomy, University College London, Gower Street, London WC1E6BT, UK}

\author{M.J. Gutmann}
\affiliation{ISIS Neutron and Muon Source, Science and Technology Facilities Council, Rutherford Appleton Laboratory, Didcot OX11 0QX, United Kingdom}

\author{S. Ricc\`{o}}
\affiliation{Department of Quantum Matter Physics, University of Geneva, 24 Quai Ernest-Ansermet, 1211 Geneva 4, Switzerland}

\author{L.S.I. Veiga}
\affiliation{London Centre for Nanotechnology and Department of Physics and Astronomy, University College London, Gower Street, London WC1E6BT, UK}

\author{C.D. Dashwood}
\affiliation{London Centre for Nanotechnology and Department of Physics and Astronomy, University College London, Gower Street, London WC1E6BT, UK}

\author{S.P. Collins}
\affiliation{Diamond Light Source Ltd., Diamond House, Harwell Science \& Innovation Campus, Didcot, Oxfordshire OX11 0DE, UK}

\author{G. Nisbet}
\affiliation{Diamond Light Source Ltd., Diamond House, Harwell Science \& Innovation Campus, Didcot, Oxfordshire OX11 0DE, UK}

\author{A. Bombardi}
\affiliation{Diamond Light Source Ltd., Diamond House, Harwell Science \& Innovation Campus, Didcot, Oxfordshire OX11 0DE, UK}
\affiliation{Clarendon Laboratory, Department of Physics, University of Oxford, Oxford OX1 3PU, England, UK}

\author{D.G. Porter}
\affiliation{Diamond Light Source Ltd., Diamond House, Harwell Science \& Innovation Campus, Didcot, Oxfordshire OX11 0DE, UK}

\author{F. Baumberger}
\affiliation{Department of Quantum Matter Physics, University of Geneva, 24 Quai Ernest-Ansermet, 1211 Geneva 4, Switzerland}
\affiliation{Swiss Light Source, Paul Scherrer Institut, CH-5232 Villigen PSI, Switzerland}

\author{A.T. Boothroyd}
\affiliation{Clarendon Laboratory, Department of Physics, University of Oxford, Oxford OX1 3PU, England, UK}

\author{D.F. McMorrow}
\affiliation{London Centre for Nanotechnology and Department of Physics and Astronomy, University College London, Gower Street, London WC1E6BT, UK}

\date{\today}

\begin{abstract}

The chemical and magnetic structures of the series of compounds Ca$_{2-x}$La$_x$RuO$_4$ [$x = 0$, $0.05(1)$, $0.07(1)$, $0.12(1)$] have been investigated using neutron diffraction and resonant elastic x-ray scattering. Upon La doping, the low temperature S-Pbca space group of the parent compound is retained in all insulating samples [$x\leq0.07(1)$], but with significant changes to the atomic positions within the unit cell. These changes can be characterised in terms of the local RuO$_6$ octahedral coordination: with increasing doping the structure, crudely speaking, evolves from an orthorhombic unit cell with compressed octahedra to a quasi-tetragonal unit cell with elongated ones. The magnetic structure on the other hand, is found to be robust, with the basic $k=(0,0,0)$, $b$-axis antiferromagnetic order of the parent compound preserved below the critical La doping concentration of $x\approx0.11$. The only effects of La doping on the magnetic structure are to suppress the A-centred mode, favouring the B mode instead, and to reduce the N\'{e}el temperature somewhat. Our results are discussed with reference to previous experimental reports on the effects of cation substitution on the $d^4$ Mott insulator Ca$_2$RuO$_4$, as well as with regard to theoretical studies on the evolution of its electronic and magnetic structure. In particular, our results rule out the presence of a proposed ferromagnetic phase, and suggest that the structural effects associated with La substitution play an important role in the physics of the system.
\end{abstract}

\maketitle

\section{Introduction}


Doped  Mott insulators host a plethora of novel electronic phases including unconventional superconductivity, pseudo-gap states, charge density wave order, etc. \cite{Lee2006} Elemental substitution offers the unique possibility to tune two of the fundamental parameters of a strongly correlated electron system, i.e. the one-electron bandwidth (\textit{bandwidth control}) and the band filling (\textit{filling control}), and thus allows a rich phase diagram of electronic and magnetic states to be accessed\cite{Imada1998a}. The doping evolution of the magnetic ground state of the parent compound, which is usually dictated by oxygen-mediated superexchange interactions, varies between different systems. In perovskite manganites the parent antiferromagnetic (AFM) order develops into a ferromagnetic (FM) phase hosting giant magnetoresistance\cite{Imada1998a}. On the other hand, in both $S=1/2$ cuprates\cite{Lee2006} and $J_\textrm{\tiny eff}=1/2$ iridates\cite{Chen2015a,gretarsson_persistent_2016,Liu2016, Pincini2017}, doping suppresses long-range AFM order leading to the formation of incommensurate spin-density wave order \cite{Cheong1991,vignolle_two_2007-1,Fujita2012,Chen2018} with a conventional paramagnetic (PM) metal eventually emerging.

Particularly interesting is the case of Mott insulators containing $d^4$ transition metal ions (such as Ru$^{4+}$, Os$^{4+}$ and Ir$^{5+}$), where moderate spin-orbit coupling (SOC) $\lambda(\mathbf{S}\cdot\mathbf{L})$ is expected to stabilize a non-magnetic $J_\textrm{\tiny eff}=0$ singlet\cite{abragam_electron_2012}. In this case, magnetic order can result from the condensation of magnetic excited states driven by intersite interactions\cite{Khomskii2014a,khaliullin_excitonic_2013}, provided that the exchange coupling is strong enough to overcome the energy of promotion of the ion to the excited state (separated from the ground state by $\lambda$). This unusual singlet magnetism has been recently predicted to give rise to a rich phase diagram as a function of electron doping\cite{chaloupka_doping-induced_2016}: here, the magnetic ground state was found to strongly depend on the relative strength of the hopping integral $t_0$, the SOC constant $\lambda$, and the correlation energy $U$. In particular, through electron doping of the parent AFM phase, the system is expected to evolve towards either a FM or a PM phase depending on the size of $\lambda$, with possibility of triplet superconductivity in the former case.

\begin{figure}
	\centering
	\includegraphics[width=0.9\linewidth]{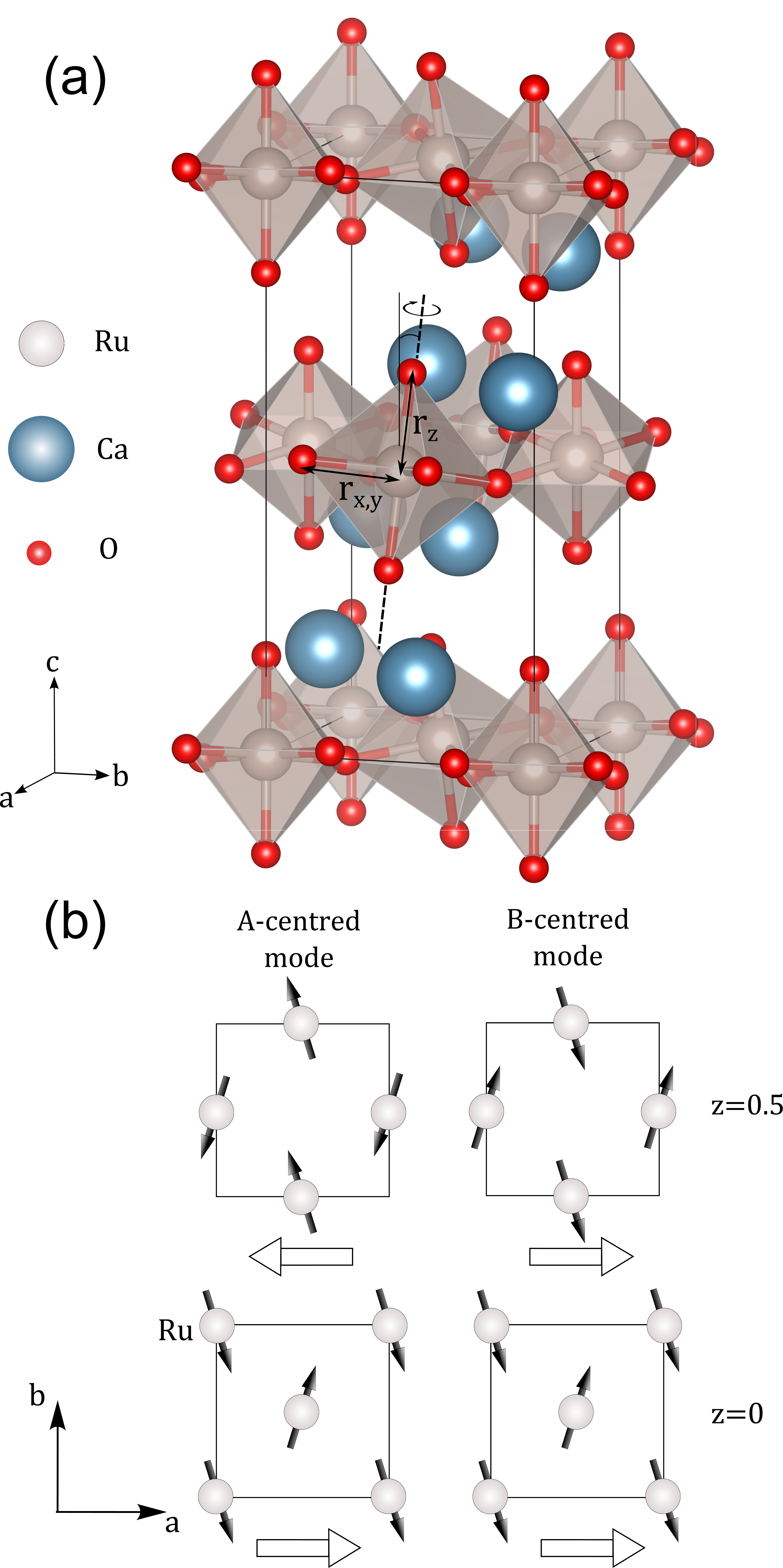}
	\caption{(Color online) (a) Ca$_2$RuO$_4$ crystal structure (Pbca space group, No. 61) highlighting the RuO$_6$ octahedra tilt and rotation discussed in the text. $r_z$ and $r_{x,y}$ correspond to the apical and in-plane Ru-O bond lengths, respectively. (b) Magnetic ordering of neighbouring RuO$_2$ layers for the A- and B-centred magnetic modes\cite{Braden1998a}. The black arrows represent the Ru$^{4+}$ ordered moments, while the white horizontal arrows  correspond to the direction of the net magnetization induced by the moment canting.}
	\label{crystal_structure}
\end{figure}

Ca$_2$RuO$_4$ is a prominent candidate for the realization of $J_{\tiny\textrm{eff}}=0$ ($L_{\tiny\textrm{eff}}=1$, $S=1$) singlet magnetism\cite{chaloupka_doping-induced_2016,khaliullin_excitonic_2013,jain_higgs_2017}. However, there is still controversy as to the precise role played by SOC in the exchange between Ru$^{4+}$ ($4d^4$) magnetic moments. Evidence supporting the proximity of Ca$_2$RuO$_4$ to a $J_{\tiny\textrm{eff}}=0$ ground state was provided by a recent inelastic neutron investigation\cite{jain_higgs_2017}, which reported the existence of the amplitude (Higgs) mode in the magnetic excitations spectrum expected to arise from the condensation of the excited $J_\textrm{\tiny eff}=1$ triplet. Other studies accounted for the spin-wave spectrum by a more conventional $S=1$ Heisenberg-like magnetic-exchange model, where a finite spin-wave gap is opened by the SOC-induced single-ion anisotropy \cite{Kunkemoller2015a,Kunkemoller2017,Zhang2017}. Neutron diffraction\cite{Braden1998a} also reveald that Ca$_2$RuO$_4$ orders at low temperature in a canted antiferromagnetic (C-AFM) structure with propagation vector $k=(0,0,0)$: the latter consists of two coexisting magnetic modes [A-centred (dominant) and B-centred], where the Ru$^{4+}$ magnetic moments predominantly lie along the elongated $\mathbf{b}$ axis of the Pbca orthorhombic lattice (see Fig.~\ref{crystal_structure}). The finite Dzyaloshinskii-Moriya interaction (DMI) (stemming from significant distortion of the RuO$_6$ octahedra\cite{Braden1998a}) causes a small net magnetization along the $\mathbf{a}$ axis, which is coupled antiferromagnetically (A-centre mode) or ferromagnetically (B-centred mode) between neighbouring RuO$_2$ layers\cite{Braden1998a}. 

\begin{figure}
	\centering
	\includegraphics[width=\linewidth]{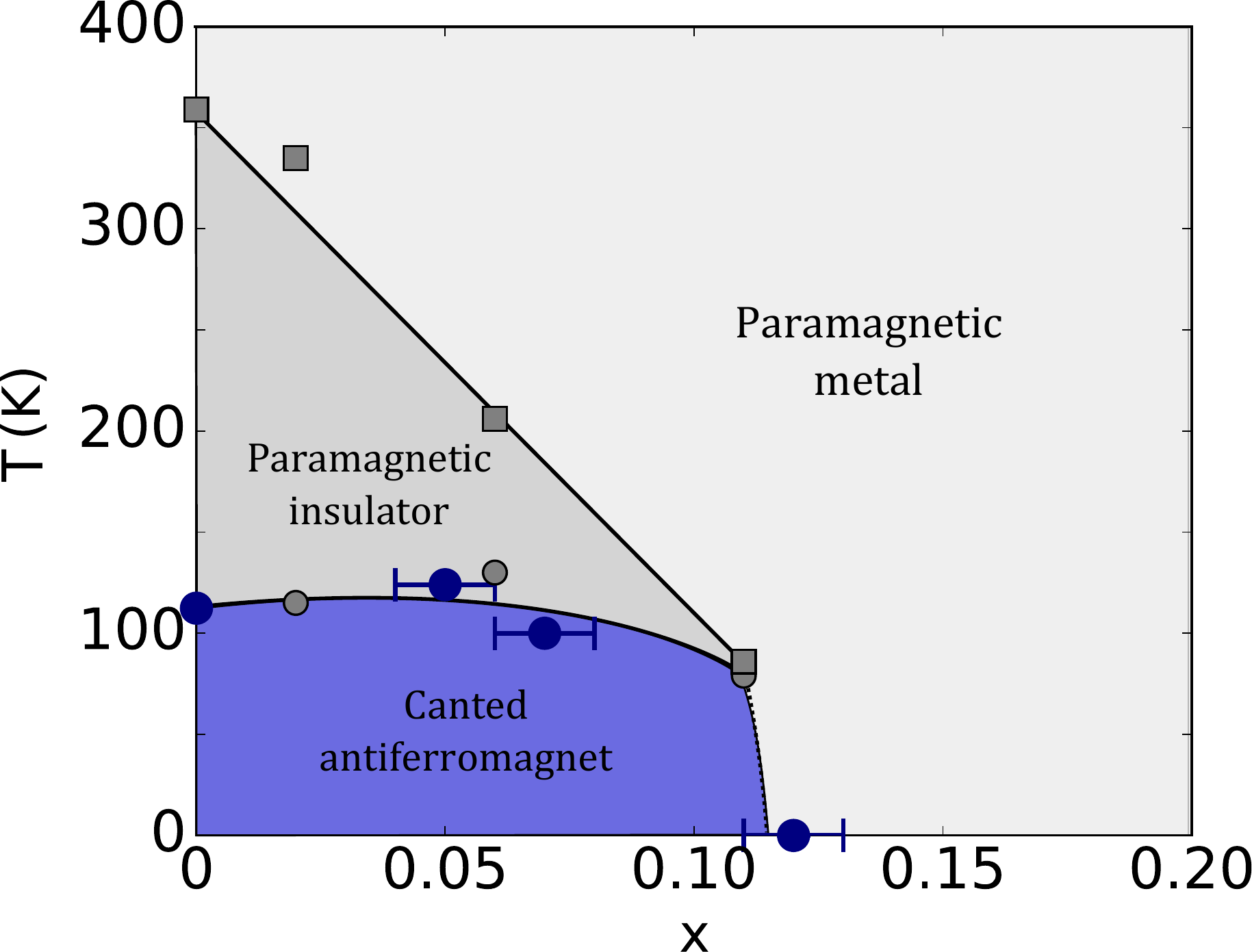}
	\caption{(Color online) Ca$_{2-x}$La$_x$RuO$_4$ temperature-doping phase diagram. The filled squares represent the temperatures (from Ref.~\onlinecite{fukazawa_filling_2001}) at which the transition between the high temperature quasi-tetragonal metallic phase (L-Pbca) and the low temperature orthorhombic one (S-Pbca) occurs. The small and large filled circles are the N\'{e}el temperatures taken from Ref.~\onlinecite{fukazawa_filling_2001} and derived from our bulk magnetization measurements\cite{supplemental_material}, respectively. The error bars reflect the uncertainty in the doping level measured by means of EDX.}
	\label{phase_diagram}
\end{figure}

The doped system Ca$_{2-x}$La$_x$RuO$_4$, where divalent Ca is replaced by trivalent La, has been recently suggested as a candidate material for the investigation of the impact of electron doping on the parent compound C-AFM order\cite{chaloupka_doping-induced_2016}. In particular, the emergence of a FM phase upon doping was predicted based on previous estimates of $\lambda$\cite{abragam_electron_2012,mizokawa_spin-orbit_2001}. Experimental studies of doped Ca$_2$RuO$_4$, however, have mainly focused on the bandwidth control achieved by substitution of divalent Sr for isovalent Ca\cite{Nakatsuji1999,Noh2005a,Wang2005a,Baier2007,Steffens2011,friedt_structural_2001}. The investigation of the magnetic properties of the La-doped compounds have been limited to bulk magnetization measurements\cite{fukazawa_filling_2001,cao_ground-state_2000,cao_ferromagnetic_2001}. The magnetization data, however, are not conclusive and have been subject to two conflicting interpretations: Fukazawa \textit{et al}\cite{fukazawa_filling_2001} interpreted the observed net magnetization as resulting from the canting of antiferromagnetically-coupled moments analogous to the parent case\cite{Braden1998a} (Fig.~\ref{phase_diagram}), while Cao \textit{et al}\cite{cao_ground-state_2000,cao_ferromagnetic_2001} proposed the presence of FM ordering, consistent with the theoretical predictions of Chaloupka and Khaliullin\cite{chaloupka_doping-induced_2016}. Despite its relevance in light of the unusual doping effects predicted for a $d^4$ Mott insulator with singlet magnetism\cite{chaloupka_doping-induced_2016}, the magnetic structure of doped Ca$_2$RuO$_4$ is thus still unresolved. 

In this paper we report on a resonant x-ray scattering (REXS) investigation of the magnetic structure of Ca$_{2-x}$La$_x$RuO$_4$ at the Ru $L_3$ and $L_2$ absorption edges. Our measurements clearly show that the AFM structure of the parent compound is retained in the doped crystals up to $x=0.07(1)$. Long-range magnetic order is destroyed at a doping level between $x=0.07(1)$ and $x=0.12(1)$, at which value the system is PM. The effect of La substitution is mainly to suppress the A-centred AFM mode of the parent compound and stabilize the B-centred one. Our findings, supported by a detailed structural characterization by means of neutron diffraction, are compared with the results observed upon Sr doping and discussed in relation to the structural changes induced by doping.

\section{Methods}

The REXS measurements were performed using the six-circle kappa diffractometer at the I16 beamline\cite{Collins2010b} of the Diamond Light Source (Didcot, UK). The scattered signal of several space-group forbidden reflections was measured tuning the incident x-rays energy to the Ru $L_3$ ($E=2.838$~keV) and $L_2$ ($E=2.967$~keV) absorption edges by means of a channel-cut Si (111) crystal. Given the relatively low photon energy, the beamline was used in a non-standard setup and extra-care was taken to minimize air absorption. The data were collected in vertical scattering geometry, using horizontal linear incident polarization (referred to as $\sigma$ polarization\cite{hill_x-ray_1996}). The samples were mounted with the $\mathbf{c}$ axis of the Pbca structure in the scattering plane for diffractometer angles set to zero; different values of the sample azimuth $\psi$ were used, where $\psi=0^{\circ}$ corresponds to having $\mathbf{b}$ in the scattering plane. Polarization analysis of the diffracted beam was achieved by means of a Pyrolytic Graphite (002) crystal in $90^{\circ}$-scattering geometry placed upstream with respect to an APD detector. This provided a scattering angle of $\theta=40.66^{\circ}$ and $\theta=38.51^{\circ}$ at the $L_3$ and $L_2$ edge, respectively. The total scattered intensity was measured using a Pilatus 100K area detector in ultra-high gain mode. The samples were cooled down below the N\'{e}el transition temperature by means of a closed-cycle cryostat. 

The crystal structure characterization was carried out by means of the Laue single-crystal diffractometer at the SXD instrument of the ISIS Neutron and Muon Source (Didcot, UK)\cite{Keen2006} while XANES spectra at the Ru $L$ edges were measured at the ID12 beamline of the European Synchrotron Radiation Facility (Grenoble, France) in total-electron yield detection mode and used to correct the REXS data for self-absorption.

\section{Sample growth and characterization}

\begin{figure}[htp]
	\centering
	\includegraphics[width=0.45\textwidth]{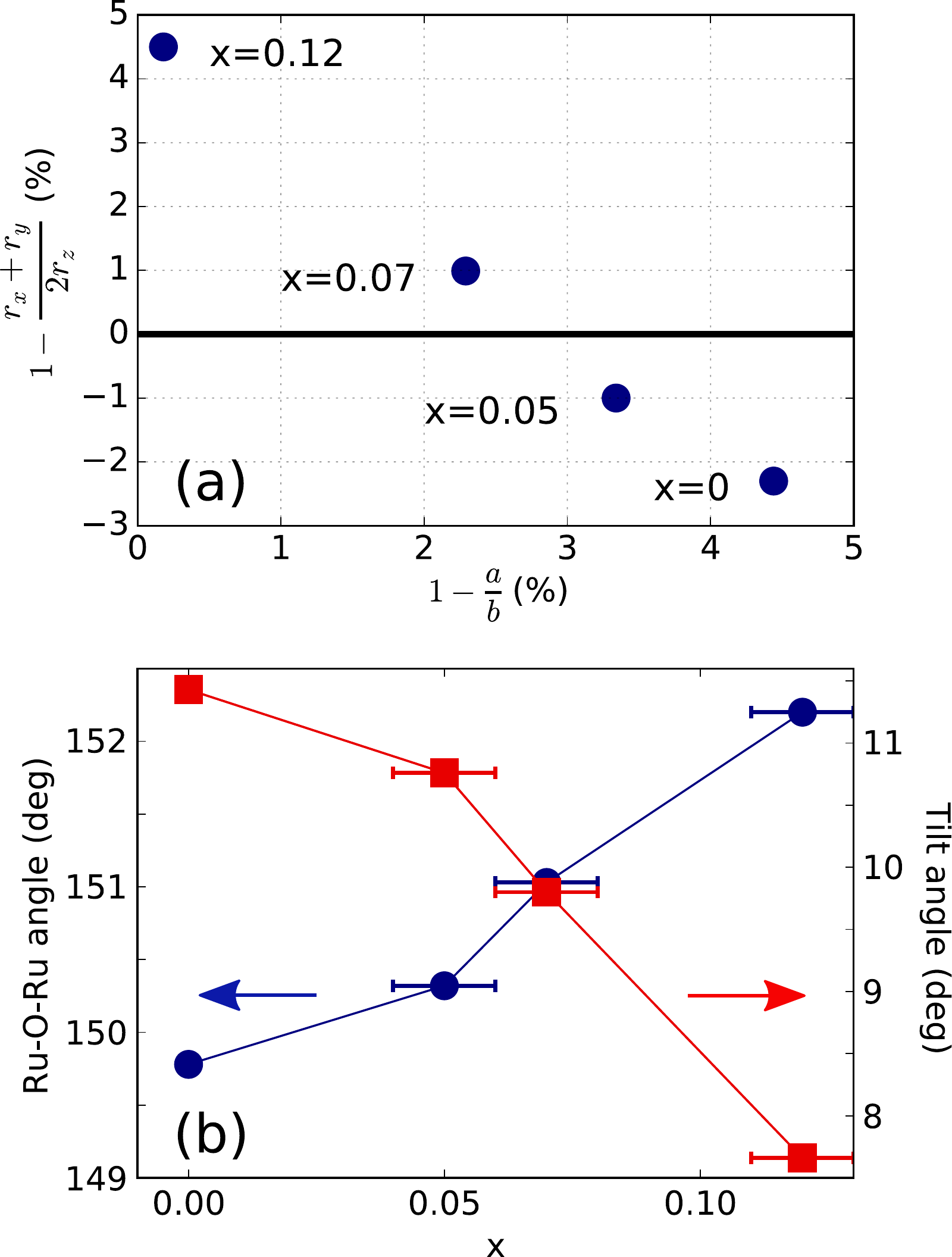}
	\caption[]{(Color online) Structural changes of Ca$_{2-x}$La$_x$RuO$_4$ as a function of doping at $T=10$~K. (a) Phase diagram as a function of the RuO$_6$ octahedra distortion $1-\frac{r_x+r_y}{2r_z}$ (with $r_{x,y}$ and $r_z$ in-plane and apical Ru-O bond length, respectively) and unit cell distortion $1-\frac{a}{b}$ ($a,b$ in-plane lattice parameters of the Pbca unit cell). The horizontal line separates the regions corresponding to octahedral compression ($1-\frac{r_x+r_y}{2r_z}<0$) and elongation ($1-\frac{r_x+r_y}{2r_z}>0$). (b) Ru-O-Ru bond angle (circles) and RuO$_6$ octahedra tilt angle away from the $\mathbf{c}$ axis (squares)  as a function of the La content. The error bars reflect the uncertainty in the doping level measured by means of EDX.}
	\label{structural_changes}
\end{figure}

\begin{figure*}[htp]
	\centering
	\includegraphics[width=\textwidth]{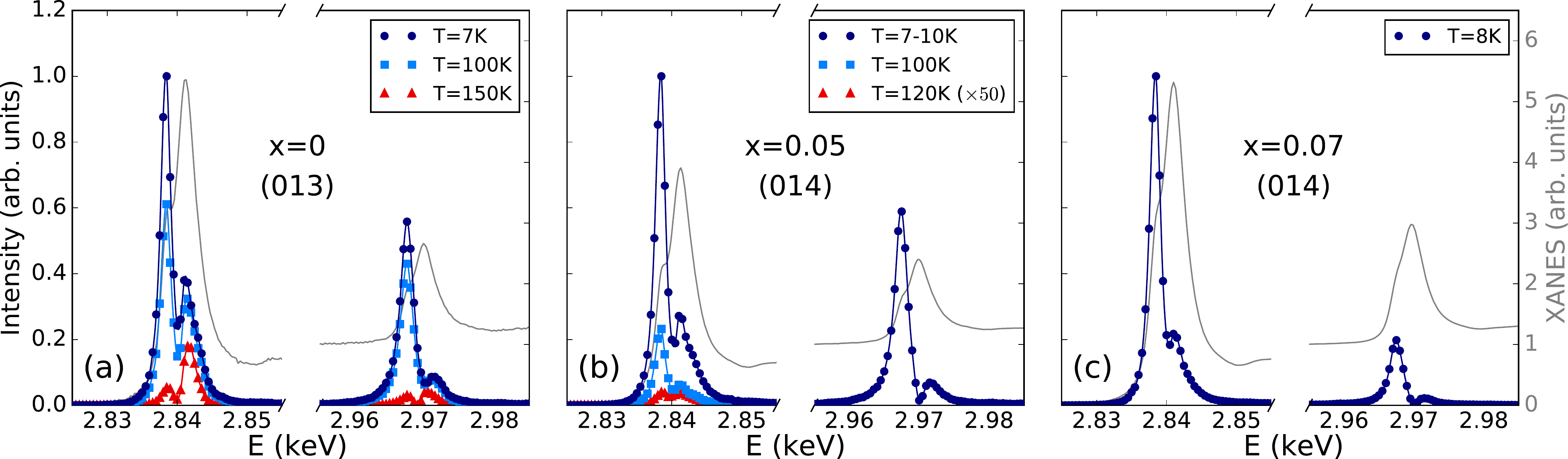}
	\caption[]{(Color online) Ru $L_3$ and $L_2$ energy resonances of magnetic diffraction peaks at different temperatures across the N\'{e}el transition in the (a) undoped, (b) $x=0.05$ and (c) $x=0.07$ sample. The filled symbols refer to the total scattered intensity corrected for self-absorption and normalized to the $L_3$ peak intensity for each sample. The solid lines represent a quadratic interpolation to the data points and are meant just as a guide to the eye. The data were measured at $\psi=0^{\circ}$  for $x=0$ and $x=0.05$ at $T=100,\,120$~K, while the low temperature $x=0.05$ and $x=0.07$ data sets correspond to an average of the spectra collected in the range $\psi=0-60^{\circ}$ and at $\psi=0^{\circ},30^{\circ}$, respectively\cite{supplemental_material}. The normalized XANES used for the self-absorption correction is also shown.}
	\label{energy_resonances}
\end{figure*}

Single crystals of Ca$_{2-x}$La$_x$RuO$_4$, with $x=0$, $0.05(1)$, $0.07(1)$ and $0.12(1)$ [corresponding to the nominal dopings $x=0$, $0.05$, $0.10$ and $0.15$, respectively], were grown through the floating zone technique using a Crystal System Corporation FZ-T10000-H-VI-VPO-IHR-PC four-mirror optical furnace. Samples were prepared in $90\%$ oxygen pressure, and the initial Ru concentration in the polycrystalline rods was about $20\%$ higher than the nominal value to compensate for evaporation during the growth. The doping level was determined by means of energy-dispersive x-ray (EDX) spectroscopy, while the bulk magnetic properties were characterized through magnetization measurements performed using a Quantum Design MPMS 3 setup\cite{supplemental_material}. The crystals used for the REXS measurements were approximately $1\times1$~mm$^2$ in size, with a crystal mosaic of about $0.05^{\circ}$ as extracted from the FWHM of the Bragg peak rocking curve. Powder samples of the same compounds were also synthesised between $1400^{\circ}$C and $1500^{\circ}$C (the temperature was increased with the La content) and 1\% O$_2$ atmosphere. The conditions were adapted from Ref.~\onlinecite{nakatsuji_ca_1997}. The parent compound ($x=0$) displays a well-documented metal to insulator transition (MIT) at $T_\textrm{\tiny MIT}=357$~K concomitant to a first-order structural transition from a high temperature quasi-tetragonal phase with a long $c$ axis (L-Pbca) to a low temperature orthorhombic one with a short $c$ axis (S-Pbca)\cite{Alexander1999,fukazawa_filling_2001,Cao1997,nakatsuji_ca_1997}. Below $T_N\approx 110$~K a phase transition to the basal plane C-AFM state also occurs\cite{Alexander1999,Braden1998a,nakatsuji_ca_1997,cao_ground-state_2000,cao_ferromagnetic_2001,fukazawa_filling_2001,Fukazawa2000}. La substitution causes the MIT and N\'{e}el temperature to decrease and be ultimately completely suppressed at a doping concentration slightly higher than $x=0.11(2)$\cite{fukazawa_filling_2001}. The corresponding temperature-doping phase diagram is shown in Fig.~\ref{phase_diagram}. Here, we report the N\'{e}el temperature derived from our bulk magnetization measurements (large filled circles with errorbars) \cite{supplemental_material} along with the results of the magnetization (small filled circles) and resistivity (filled squares) data collected by Fukazawa \textit{et al}\cite{fukazawa_filling_2001}.

\section{Results}

\subsection{Structural characterization}

\begin{figure}[htp]
	\centering
	\includegraphics[width=0.45\textwidth]{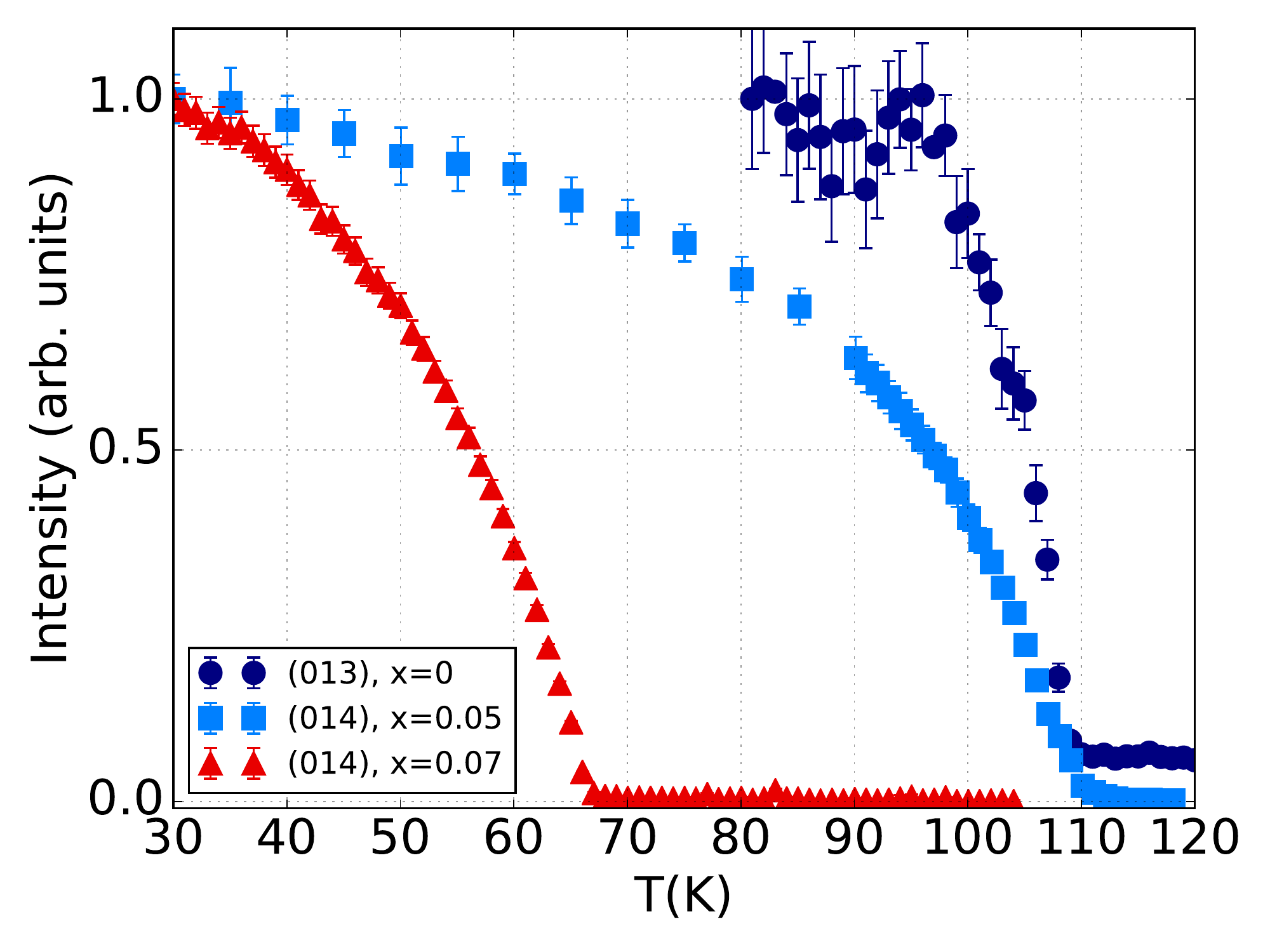}
	\caption[]{(Color online) Temperature dependence of the $(013)$ (dark blue circles) and $(014)$ magnetic diffraction peaks in the undoped and $x=0.05$ (light blue squares) and $x=0.07$ (red triangles) samples, respectively. The data points correspond to the total diffracted intensity integrated over a rocking curve at $E=2.838$~keV and $\psi=0^{\circ}$ and normalized to the low temperature value.}
	\label{temperature_dependence}
\end{figure}

The structural properties of Ca$_{2-x}$La$_x$RuO$_4$ as a function of the doping level were investigated by means of single crystal neutron diffraction at $T=10$~K. The doped samples retain the same space group (Pbca, No.61) of the parent compound. However, the different Shannon radii of the La$^{3+}$ ($r=1.22$~\AA) and Ca$^{2+}$ ($r=1.18$~\AA) ions\cite{Shannon1976,fukazawa_filling_2001} result in significant structural changes. The unit cell of Ca$_2$RuO$_4$ is shown in Fig.~\ref{crystal_structure}(a) along with the definition of relevant structural parameters, while the main results of the neutron data refinement are summarized in Fig.~\ref{structural_changes}. The structural changes can be described in terms of the following four parameters: (i) unit cell distortion $ 1-\frac{a}{b}$, where $a$ and $b$ are the in-plane lattice constants of the Pbca unit cell; (ii) octahedral distortion $1-\frac{r_x+r_y}{2r_z}$, where  $r_{x,y}$ and $r_z$  are the in-plane and apical Ru-O bond lengths of the RuO$_6$ octahedra, respectively; (iii) Ru-O-Ru bond angle; (iv) octahedral tilt angle away from the crystallographic $\mathbf{c}$ axis. The unit cell of the parent compound at low temperature is orthorhombic, with a $b$ lattice parameter elongated by $4.4\%$ with respect to $a$. Large distortions away from the perfect perovskite structure ($1-\frac{r_x+r_y}{2r_z}=0$, Ru-O-Ru angle~$=180^{\circ}$ and tilt angle~$=0^{\circ}$) are also present: the RuO$_6$ octahedra are significantly compressed along the local $z$ axis ($1-\frac{r_x+r_y}{2r_z}<0$) and display both a sizeable rotation around the apical Ru-O bond direction and tilt away from the $\mathbf{c}$ axis [Fig.~\ref{crystal_structure}(a)]. La substitution is found to cause a reduction of the orthorhombicity of the unit cell ($1-\frac{a}{b}\rightarrow 0$), in agreement with a previous study\cite{fukazawa_filling_2001}, and an elongation of the octahedral cage ($1-\frac{r_x+r_y}{2r_z}>0$): this results in the phase diagram of Fig.~\ref{structural_changes}(a), where, for increasing doping levels, the system evolves from an orthorhombic cell with compressed octahedra ($x=0,\,0.05$) to a quasi-tetragonal cell with elongated ones ($x=0.07,\,0.12$). The rotations of the octahedral cage are also reduced [Fig.~\ref{structural_changes}(b)] and the structure tends to relax towards the undistorted perovskite lattice. Further details on the structural refinement and additional room temperature results are reported in the Supplemental Material\cite{supplemental_material} and Ref.~\onlinecite{Ricco2018}, which the reader is referred to for the corresponding CIF files.


\bgroup
\footnotesize
\def\arraystretch{1.2}
\begin{table}[htp]
	\centering
	
	\resizebox{0.45\textwidth}{!}{
		\begin{tabular}{ccllll}
			\hline\hline
			& \multicolumn{5}{c}{AFM reflections}                             \\ \hline
			A-centred mode & \multicolumn{5}{c}{$(100)$, $(011)$, $(013)$, $(120)$}          \\
			B-centred mode & \multicolumn{5}{c}{$(010)$, $(101)$, $(012)$, $(103)$, $(014)$} \\ \hline\hline
		\end{tabular}
		
	}
	\caption{Space-group forbidden magnetic reflections arising from the main AFM order of the two magnetic modes in Ca$_2$RuO$_4$ \cite{Braden1998a,Bombardi_undoped}.}
	\label{tab:AFM_reflections}
\end{table}
\egroup

\subsection{Magnetic structure}

\begin{figure*}[htp]
	\centering
	\includegraphics[width=0.9\textwidth]{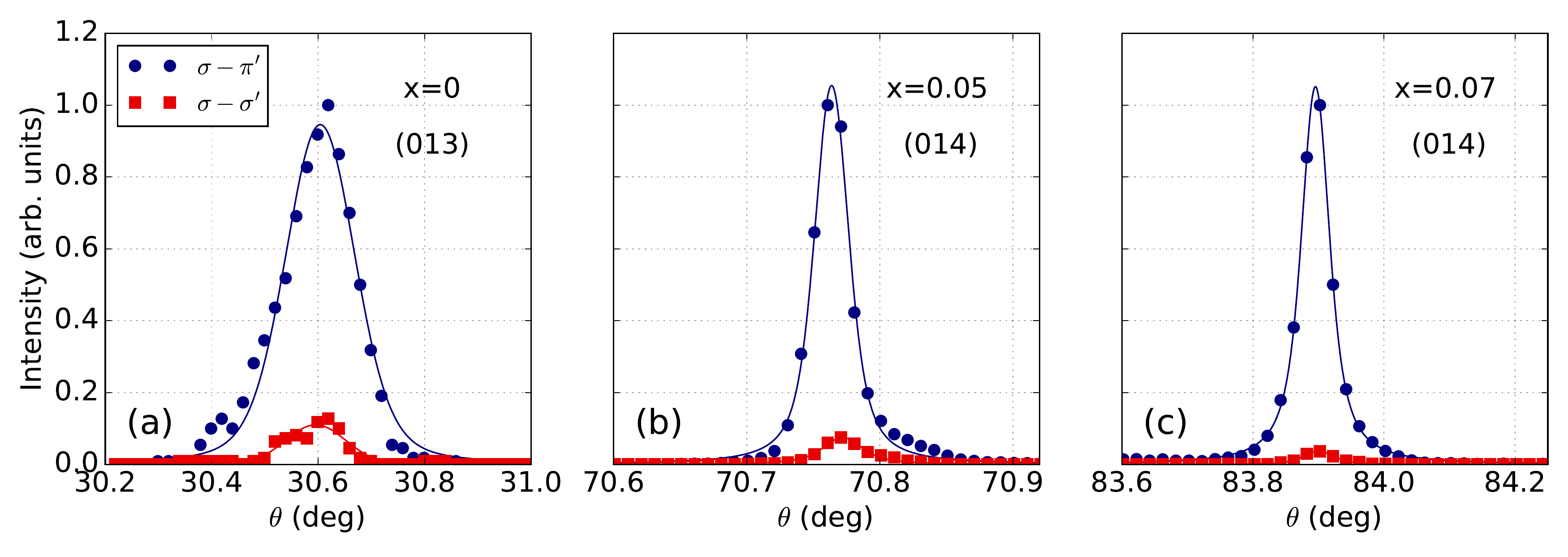}
	\caption[]{(Color online) Polarization dependence of magnetic diffraction peaks in the (a) undoped, (b) $x=0.05$ and (c) $x=0.07$ sample at $\textrm{T}=7-8$~K. The filled symbols refer to the scattered intensity measured over a rocking scan in the $\sigma-\pi'$ (blue circles) and $\sigma-\sigma'$ (red squares) channels of the polarization analyser crystal normalized to the $\sigma-\pi'$ peak intensity, while the solid lines represent a fit to Voigt profile. The data were collected at $E=2.838$~keV and $\psi=110^{\circ},\,50^{\circ}\,30^{\circ}$ for $x=0,\,0.05,\,0.07$, respectively.}
	\label{polarization_dependence}
\end{figure*}

The globally AFM A-centred and weakly FM B-centred modes [see Fig.~\ref{crystal_structure}(b)] give rise to two separate sets of AFM space-group forbidden reflections (Table~\ref{tab:AFM_reflections}) for moments along the $\mathbf{b}$ axis\cite{Braden1998a,Bombardi_undoped}, which can be selectively accessed in a scattering experiment. The AFM coupling of the net magnetization induced by the moment canting between consecutive RuO$_2$ layers in the A-centred structure [Fig.~\ref{crystal_structure}(b)] results in additional weak magnetic reflections (not listed in Table~\ref{tab:AFM_reflections}): the signal at the corresponding $(hkl)$ values, however, was found \cite{supplemental_material} to be dominated by anisotropic tensor of susceptibility (ATS) scattering\cite{Dmitrienko1983}. In contrast to the previous REXS study on the parent compound\cite{zegkinoglou_orbital_2005}, where the $(100)$ magnetic reflection was probed, our scattering geometry limited our investigation to the magnetic diffraction peaks with a non-zero $l$ component [i.e. of the type $(h0l)$ or $(0kl)$]. 

\begin{figure}[htp]
	\centering
	\includegraphics[width=0.4\textwidth]{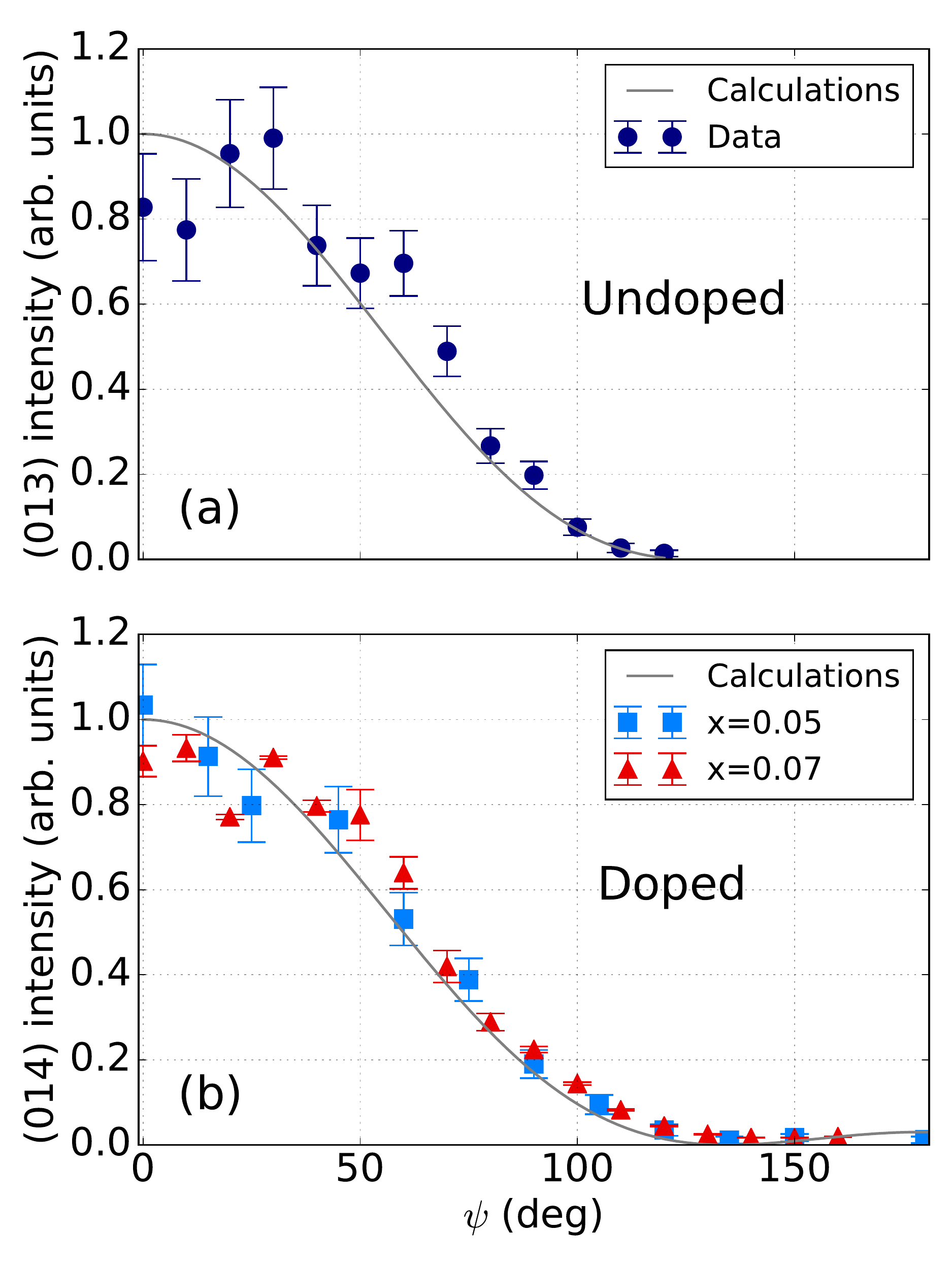}
	\caption[]{(Color online) Azimuthal dependence [azimuthal reference $(010)$] of the $(013)$ and $(014)$ magnetic diffraction peaks in the  (a) undoped and  (b) doped ($x=0.05,\,0.07$) samples at $\textrm{T}=7-8$~K, respectively. The filled symbols refer to the scattered intensity integrated over a rocking curve at $E=2.838$~keV corrected for the geometry-dependent self-absorption factor\cite{supplemental_material}. The solid lines represent the calculated azimuthal dependence (except for an arbitrary scale factor) assuming the C-AFM structure with $\mathbf{b}$ axis moments reported by Braden \textit{et al}\cite{Braden1998a}. The intensity is normalized to the calculated value at $\psi=0^{\circ}$.}
	\label{azimuthal_dependence}
\end{figure}

The energy dependence of the $(013)$ magnetic peak (A-centred mode) in the parent compound is reported in Fig.~\ref{energy_resonances}(a). The data have been corrected for self-absorption using the corresponding XANES signal\cite{supplemental_material} [also shown in Fig.~\ref{energy_resonances}(a)]. As already reported by Zegkinoglou \textit{et al}\cite{zegkinoglou_orbital_2005}, a strong resonant enhancement of the diffracted intensity is present at both the Ru $L_3$ and $L_2$ absorption edges. The resonance originates from electric dipole $2p\rightarrow 4d$ transitions which directly probe the partially filled Ru $4d$ states responsible for magnetism. Each resonance displays two distinct features residing at $E=2.8383(2),\,2.8417(2)$~keV ($L_3$) and $E=2.9674(2),\,2.9721(2)$~keV ($L_2$), which arise from transitions to the crystal-field-split $t_{2g}$ and $e_g$ Ru $4d$ orbitals, respectively\cite{zegkinoglou_orbital_2005}. The $(013)$ signal is largely magnetic in origin as demonstrated by its temperature (Fig.~\ref{temperature_dependence}), polarization [Fig.~\ref{polarization_dependence}(a)] and azimuthal dependence [Fig.~\ref{azimuthal_dependence}(a)]. A small contribution from ATS scattering might also be present\cite{supplemental_material}, as evidenced by the weak diffracted signal observed above the N\'{e}el temperature (see blue circles in Fig.~\ref{temperature_dependence}) and the residual intensity in the $\sigma-\sigma'$ polarization channel [Fig.~\ref{polarization_dependence}(a)]. The latter is also partially accounted for by the leakage from the $\sigma-\pi'$ channel, caused by the fact that the scattering angle of the analyser crystal was not exactly $90^{\circ}$.

\begin{figure}[htp]
	\centering
	\includegraphics[width=0.4\textwidth]{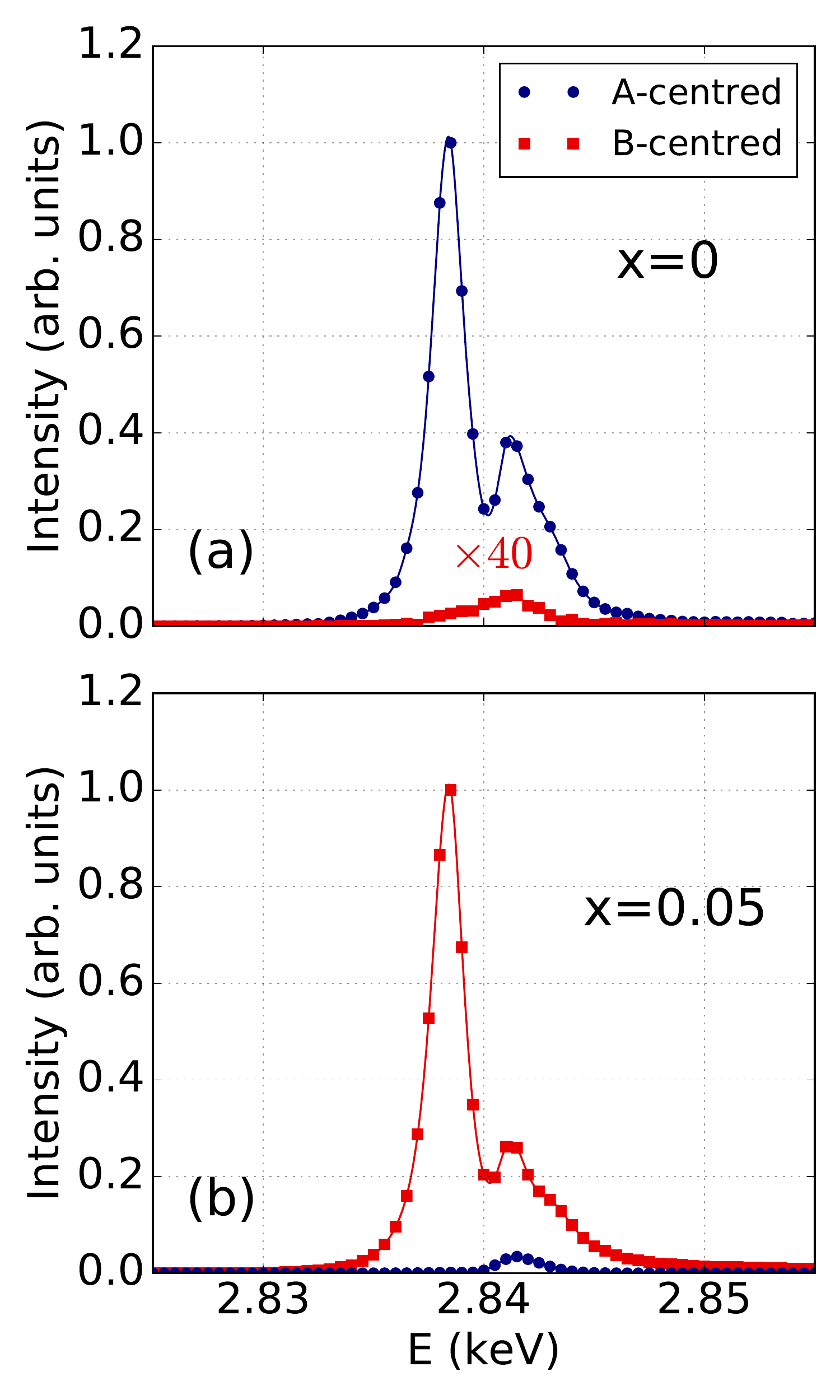}
	\caption[]{(Color online) Comparison between the $(013)$ (blue circles) and $(014)$ (red squares) Ru $L_3$ resonances in the (a) undoped and (b) $x=0.05$ sample. The filled symbols refer to the total scattered intensity at $\textrm{T}=7-8$~K corrected for self-absorption and normalized to the peak intensity of the dominant mode. The solid lines represent a quadratic interpolation to the data points and are meant just as a guide to the eye. The data were measured at $\psi=0^{\circ}$, apart from the $(014)$ in the undoped sample, for which $\psi=80^{\circ}$. }
	\label{mode_AvsB}
\end{figure}

As well as the $(013)$, a large energy resonance was also found for the $(011)$ A-centred reflection\cite{supplemental_material}.  Several B-centred peaks (Table~\ref{tab:AFM_reflections}) were also investigated at various sample azimuth values, but no significant diffracted intensity was detected in the parent compound. This is clearly shown in Fig.~\ref{mode_AvsB}(a), where the $(013)$ and $(014)$ self-absorption corrected Ru $L_3$ resonances at low temperature are reported on the same scale. The $(014)$ intensity is negligible and mostly resonates at the $e_g$ levels energy, thus suggesting that the signal is dominated by weak ATS scattering. This is consistent with previous neutron scattering measurements\cite{Braden1998a}, which found a prevalence of the A-centred mode.

In order to verify whether the magnetic structure of the parent compound is retained upon La substitution, we probed several A-centred and B-centred magnetic reflections in the doped samples. A significant diffracted intensity was found at the same $k=(0,0,0)$ reflections. However, in stark contrast to the undoped crystal, a large resonance is present for the B-centred peaks only [Fig.~\ref{mode_AvsB}(b)]. The B-centred resonances display similar features to the A-centred ones in the parent compound, as illustrated in Fig.~\ref{energy_resonances}(b-c) for the $(014)$ reflection. In particular, a double $t_{2g}-e_g$ peak with a comparable resonant enhancement is present at both the Ru $L_3$ and $L_2$ edge. A minor exception is represented by the weaker $L_2$ $(014)$ resonance in the $x=0.07$ sample, which could be attributed to changes in the Ru$^{4+}$ electronic levels induced by doping (see Sec.~\ref{sec:discussion}). It should be noticed, however, that the resonant enhancement of the $(103)$ magnetic reflection at the two edges is similar to the one observed at lower La content\cite{supplemental_material}: therefore, a definitive conclusion on the impact of doping on the Ru$^{4+}$ electronic structure cannot be drawn from the data of Fig.~\ref{energy_resonances}.

\begin{figure}[htp]
	\centering
	\includegraphics[width=0.49\textwidth]{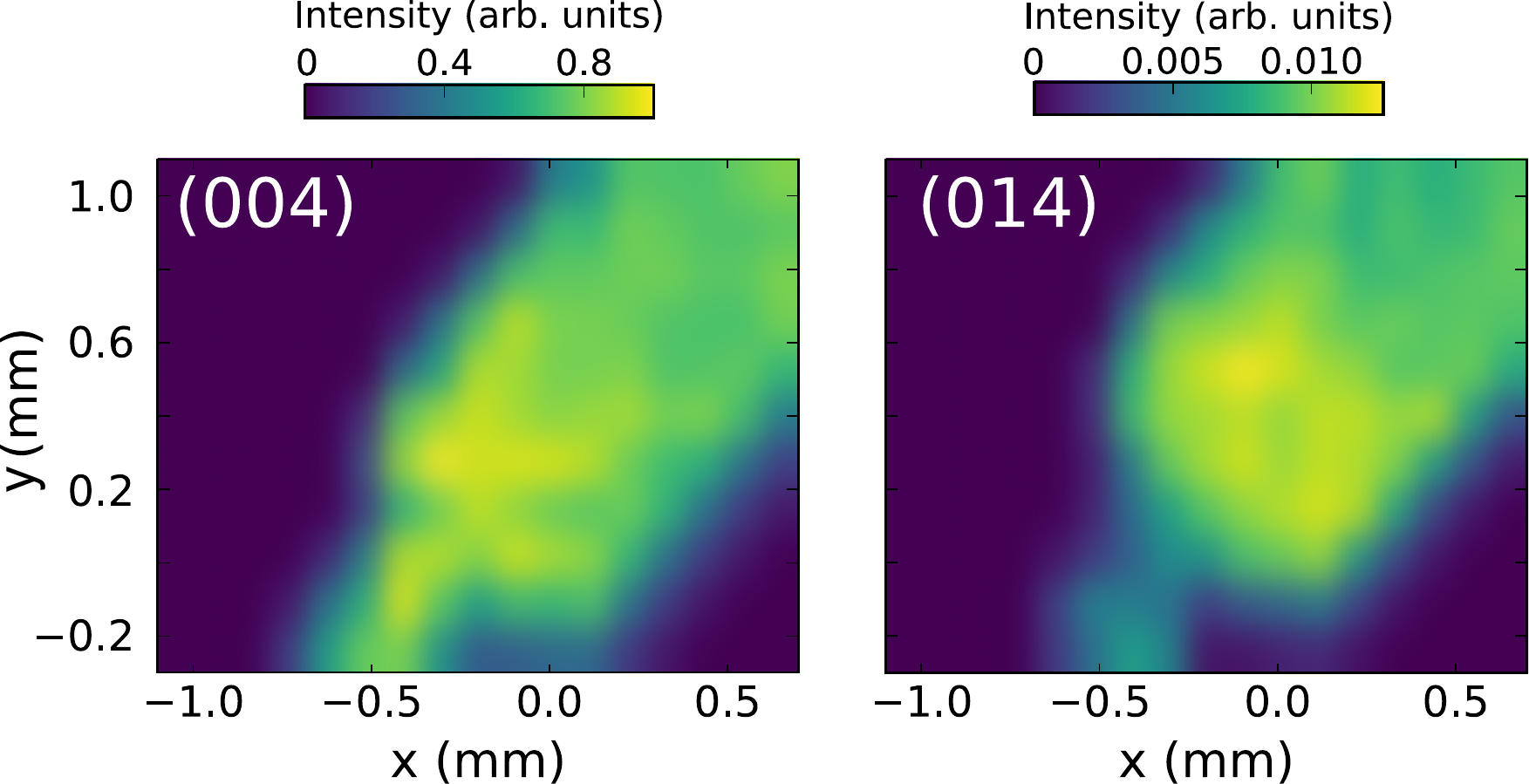}
	\caption[]{(Color online) Intensity maps of the $(004)$ Bragg peak and $(014)$ magnetic reflection ($T=8$~K, $\psi=0^{\circ}$) in the $x=0.05$ sample as a function of the incident x-ray beam position on the crystal (beam size $0.1\times0.1$~mm$^2$). The color scale represents the integrated intensity over a rocking scan measured in each position normalized to the $(004)$ peak value (the dark regions are off the sample).}
	\label{(014)_map}
\end{figure}

The B-centred mode was found to be predominant across the whole crystal without any significant spatial dependence. This is shown in the color maps of Fig.~\ref{(014)_map}, where the spatial dependence of the magnetic $(014)$ diffraction peak in the $x=0.05$ crystal is reported along with the one of the $(004)$ Bragg peak. The measurements were performed translating the sample under the beam focal spot and collecting a rocking scan in each position. The size of the incident beam was reduced to $0.1\times0.1$~mm$^2$ through a set of slits. The $(014)$ intensity shows limited variations throughout the measured sample area, with minor changes resulting from trivial inhomogeneities in the crystal quality across the sample [as can be seen from the comparison to the $(004)$ map]. This excludes the presence of phase-separated domains of prevalent A or B character.

In addition to the resonant enhancement, the magnetic origin of the $k=(0,0,0)$ space-group forbidden reflections in the doped samples is strongly supported by the fact that (i) the scattered intensity vanishes upon warming beyond $T_N\approx110$~K ($T_N\approx70$~K) in the $x=0.05$ ($x=0.07$) sample and (ii) the scattered signal is predominantly $\pi'$ polarized [Fig.~\ref{polarization_dependence}(b-c)] regardless of the x-ray energy and $\psi$ value chosen for the measurements\cite{supplemental_material}, as expected for dipole resonant magnetic scattering\cite{hill_x-ray_1996}. As for the parent compound, the weak intensity in the $\sigma-\sigma'$ polarization channel is due to leakage from the $\sigma-\pi'$ channel and $\sigma'$-polarized ATS scattering. The azimuthal dependence of the $(014)$ [Fig.~\ref{azimuthal_dependence}(b)] and $(103)$ (see Supplemental Material\cite{supplemental_material}) reflections is also consistent with the calculations performed assuming the C-AFM structure with $\mathbf{b}$ axis moments reported for pure Ca$_2$RuO$_4$\cite{Braden1998a}. 

\section{Discussion}\label{sec:discussion}

Our results clearly show that the doped samples retain the same $k=(0,0,0)$ C-AFM structure of the parent compound. In particular, as shown by the azimuthal dependence of Fig.~\ref{azimuthal_dependence}, the Ru$^{4+}$ moments preserve their $\mathbf{b}$ alignment. The impact of La substitution is limited to a suppression of the globally AFM A-centred mode, predominant in the parent compound, and a concomitant stabilisation of the B-centred structure, where a weak net magnetization is present as a result of the FM alignment of the canting-induced net moments. A similar effect has been reported both in Ca$_{2-x}$Sr$_x$RuO$_4$\cite{friedt_structural_2001} and Ca$_2$RuO$_4$ under pressure\cite{Steffens2005}. Contrary to the La case, Sr and Ca are both divalent: Sr doping thus only realizes a bandwidth control of the parent insulator, due to the different Shannon radii of the Sr$^{2+}$ ($r=1.31$~\AA) and Ca$^{2+}$ ($r=1.18$~\AA) ions\cite{Shannon1976}. Moreover, the internal chemical pressure originating from Sr doping and the application of external pressure have similar structural effects\cite{Steffens2005}. The latter also resemble our neutron scattering results and therefore suggest that the changes in the crystal structure may be responsible for the observed transition from A- to B-type order. In this respect, the evolution of the Ru-O-Ru bond angle upon doping could be of particular importance since it directly controls the oxygen-mediated superexchange between Ru atoms\cite{goodenough1963magnetism}. This could favour the FM alignment of the net moments and might be one of the mechanisms involved in the stabilisation of the B-centred structure. 

The ground-state wave function of the Ru$^{4+}$ ion was described in terms of an admixture of $xy$, $yz$ and $zx$ orbitals, whose respective contribution depends on the relative strength of the tetragonal crystal field $\delta$ and SOC $\lambda$\cite{mizokawa_spin-orbit_2001,fatuzzo_spin-orbit-induced_2015,Nakatsuji2000d,Liu2011a,Liu2013,Kurokawa2002,Das2018}. Within this picture, the tuning of the crystal field caused by the elongation of the RuO$_6$ octahedra [Fig.~\ref{structural_changes}(a)] is expected to result in an enhanced $xy$ hole occupancy in the doped compounds with respect to the parent case. This is indeed what was reported for Ca$_{2-x}$Sr$_x$RuO$_4$: here, the orbital degeneracy control achieved by Sr doping shifts the Ru $4d$ $xy$ bands towards the Fermi level and turns the AFM exchange of the parent compound into a FM one at $x_c=0.5$ \cite{Nakatsuji2000d,Fang2004a}. A similar effect might be present also in the case of La doping and contribute to the stabilisation of the B-centred structure. Moreover, Hartree-Fock calculations \cite{mizokawa_spin-orbit_2001,Kurokawa2002} predicted that the increase in the $xy$ holes population driven by tetragonal elongation should result into a $z$ alignment of the Ru magnetic moments. Although the presence of a finite $z$ component cannot be excluded by our measurements, a spin-flop transition to a $c$-axis AFM structure in the tetragonally elongated $x=0.07$ sample is ruled out by the azimuthal dependence of the scattered intensity [Fig.~\ref{azimuthal_dependence}(b)].

The evolution from A- to B-type magnetic order naturally explains the FM behaviour seen in bulk magnetization measurements\cite{cao_ground-state_2000,cao_ferromagnetic_2001,fukazawa_filling_2001} as resulting from the weak FM component of the B-centred C-AFM mode, rather than a doping-induced local FM alignment (FM polarons) of Ru moments\cite{cao_ground-state_2000,cao_ferromagnetic_2001}. This is consistent with the interpretation given by Fukazawa \textit{et al}\cite{fukazawa_filling_2001} based on their SQUID data. Given their sensitivity to the macroscopic magnetization of the sample, bulk measurements alone are not capable of unequivocally distinguishing between the two scenarios: similar results are indeed expected regardless of whether the magnetization arises from a FM ordering with a small ordered moment or an AFM structure with a weak net magnetization due to moments canting. A further confirmation comes from the measurements performed on non-stoichiometric Ca$_2$RuO$_{4+\delta}$\cite{Braden1998a}: here, a B-type C-AFM structure was also found and the bulk magnetization shows similar features to the ones of the La-doped compounds. The increase in the net moment seen for increasing levels of La content\cite{cao_ground-state_2000} is also compatible with the C-AFM scenario, where it could arise from an increase of either the B-centred mode volume fraction or the DMI-induced canting angle. However, minor changes in the magnitude of the latter could not be investigated since the resulting weak FM component, contrary to the case of the A-centred mode, does not give rise to any space-group forbidden reflections.

Although most of the literature on Ca$_{2-x}$La$_x$RuO$_4$ focuses on the filling control of the Ru $4d$ bands associated with the extra electron introduced by the La$^{3+}$ ion\cite{chaloupka_doping-induced_2016,cao_ground-state_2000,cao_ferromagnetic_2001,fukazawa_filling_2001}, our findings show that the concomitant structural effects are likely to play a crucial role in the physics of the system. This scenario is somewhat confirmed by recent resistivity and ARPES measurements on Pr-doped Ca$_2$RuO$_4$\cite{Ricco2018}, which suggest that, in contrast to lightly doped cuprates \cite{Damascelli2003} and iridates \cite{DeLaTorre2015, kim_fermi_2014}, the doped electrons remain fully localized in the S-Pbca phase irrespective of the Pr content. A pronounced sensitivity to the doping-induced structural changes is also expected in light of the importance of lattice energies in the stabilization of the low-temperature insulating state highlighted by recent \textit{ab-initio} calculations in the parent compound \cite{Han2018}


Nonetheless, Ca$_{2-x}$La$_x$RuO$_4$ was explicitly reported as an example of an electron-doped system in recent theoretical calculations\cite{chaloupka_doping-induced_2016}. The latter found that the injection of free carriers on the ground state of $d^4$ Mott insulators dramatically depends on the interplay between the exchange interaction $K$, the hopping integral $t_0$ and the SOC constant $\lambda$. In particular, starting from the parent compound AFM ground state, a FM phase is predicted to rapidly appear upon electron doping for sufficiently weak SOC. This scenario has been explicitly supported by Chaloupka and Khaliullin\cite{chaloupka_doping-induced_2016} who, backed by the interpretation of the bulk magnetization data given by Cao \textit{et al}\cite{cao_ground-state_2000}, based their conclusion on previous estimates for $\lambda$\cite{mizokawa_spin-orbit_2001,abragam_electron_2012}.  Our measurements, however, show that the FM phase is not realized in Ca$_{2-x}$La$_x$RuO$_4$. AFM order survives up to $x=0.07(1)$, while the system is found to be PM at $x=0.12(1)$. Assuming the filling of the Ru $4d$ bands is the dominant effect in the physics of La-doped Ca$_2$RuO$_4$, this sets a lower boundary to the SOC constant: considering $t\approx300$~meV\cite{chaloupka_doping-induced_2016}, the FM phase is predicted to be absent for $\lambda>77$~meV. The latter estimate is compatible with the value $\lambda\approx200$~meV from a recent O $K$-edge RIXS investigation\cite{fatuzzo_spin-orbit-induced_2015}. Approximating to $x=0.10$ the value of the doping level at which the transition between the C-AFM and PM states occurs\cite{fukazawa_filling_2001}, the phase diagram of Ref.~\onlinecite{chaloupka_doping-induced_2016} gives $\lambda\approx 400$~meV. This is comparable to the value found for $5d$ transition metal oxides\cite{kim_novel_2008} and thus seems overestimated. The theoretical phase diagram of Ref.~\onlinecite{chaloupka_doping-induced_2016}, however, neglects the structural changes discussed in the present investigation (as well as distortions away from the perfect cubic symmetry of the RuO$_6$ octahedra and deviations from two-dimensionality) and, as a result, it does not provide with an accurate description of the physics of Ca$_{2-x}$La$_x$RuO$_4$.

\section{Concluding remarks}

In conclusion, our REXS investigation establishes the persistence of the C-AFM structure of the parent compound in Ca$_{2-x}$La$_x$RuO$_4$. The AFM order was found to be present up to $x=0.07(1)$, while long-range order is absent in the $x=0.12(1)$ sample. La substitution suppresses the globally AFM A-centred mode, dominant in pure Ca$_2$RuO$_4$, and favours the B-centred structure, which displays a weak net magnetization as a result of the moments canting. The latter naturally explains the net magnetization observed in previously published bulk magnetization measurements on the doped samples. Our results rule out the presence of a FM phase and also suggest that the structural changes which accompany La doping are likely to play a pivotal role in the observed magnetic properties and should be considered alongside the electron doping effect in any meaningful description of the physics of the system.

\begin{acknowledgments}
The authors would like to thank the I10 beamline staff at DLS for their help during the bulk magnetization measurements. We acknowledge Diamond Light Source for time on beamline I16 under Proposal No. MT-15323, MT-15952, MT-15867 and MT-18934. Experiments at the ISIS Neutron and Muon Source were supported by a beamtime allocation from the Science and Technology Facilities Council. This work is supported by the UK Engineering and Physical Sciences Research Council (Grants No. EP/N027671/1 and No. EP/N034694/1) and by the Swiss National Science Foundation (Grant No. 200021\_153405).
\end{acknowledgments}

\bibliography{Bibliography}

\end{document}